\documentclass[aps,twocolumn,prb]{revtex4}
\usepackage{amsmath}
\usepackage{graphicx}

\begin{document}

\title[Spin Hall Effect]{Suppression of the Persistent Spin Hall Current by
Defect Scattering}
\author{Jun-ichiro Inoue}
\email{inoue@nuap.nagoya-u.ac.jp}
\affiliation{Department of Applied Physics, Nagoya University, Nagoya 464-8603, Japan,
and CREST-JST, Japan}
\author{Gerrit E.W. Bauer}
\affiliation{Department of NanoScience, Delft University of Technology, Lorentzweg 1,
2628CJ Delft, The Netherlands}
\author{Laurens W. Molenkamp}
\affiliation{Physikalisches Institut (EP3), Universit\"{a}t W\"{u}rzburg, D-97074 W\"{u}%
rzburg, Germany}
\pacs{PACS number: }

\begin{abstract}
We study the linear response spin Hall conductivity of a two-dimensional
electron gas (2DEG) in the presence of the\ Rashba spin orbit interaction in
the diffusive transport regime. When defect scattering is modeled by
isotropic short-range potential scatterers the spin Hall conductivity
vanishes due to the vertex correction. A non-vanishing spin Hall effect may
be recovered for dominantly forward defect scattering.
\end{abstract}

\maketitle


Spintronics is the rapidly developing field of research aiming to use not
only the charge but also the spin degree of freedom of electrons in
electronic circuits and devices.\cite{wolf} In order to be compatible with
microelectronic technology, effective spin injection into conventional
semiconductors is necessary. Injection of spins via attached ferromagnets
has turned out to be quite difficult.\cite{fiederling,ohno} This is one
motivation to investigate the possibilities to make use of the spin-orbit
(SO) interaction, which may spin-polarize a non-magnetic conductor simply by
applying a source-drain bias.\cite%
{vasko,levitov,edelstein,inoue,murakami,sinova} The two-dimensional electron
gas (2DEG) is an ideal model system to investigate the physics of these
effects. In sufficiently asymmetric confinement potentials the so-called
Rashba term dominates the SO interaction.\cite{rashba} The Datta spin
transistor concept \cite{datta} is based on the tunability of the Rashba
interaction by an external gate potential.\cite{nitta}

Applying an electric field in the $x-$direction of a Rashba 2DEG spanning
the $x,y-$plane induces a charge current in the $x-$direction, but also a
homogeneous spin accumulation in the $y-$direction proportional to the field
strength.\cite{edelstein,inoue} Recently, Sinova \textit{et al}. reported a
persistent spin Hall current \cite{sinova} for a ballistic Rashba 2DEG. The
acceleration of the electrons by the external electric field (along the $x$%
-direction) modifies the SO-induced pseudo magnetic field such that spin are
tilted out of the 2DEG plane in directions that are opposite for positive
and negative lateral momentum ($k_{y}$) states. This corresponds to a flow
of $s_{z}=+1/2$ and $-1/2$ spins in opposite directions without a
corresponding net charge transport.\cite{sinova} The authors suggest that
the spin Hall current should be rather robust against disorder scattering,
which implies that the effect is measurable in Hall bars of mesoscopic
dimensions. Note that the ballistic spin Hall effect is quite different from
the spin Hall effect reported earlier for diffuse paramagnetic metals, which
is caused essentially by impurity scattering.\cite{hirsch,zhang} In the 
\emph{weak scattering} regime, in which the broadening is smaller than the
SO-induced splitting of the energy bands, the life-time broadening of the
self-energy has recently been found to have vanishing effect on the
ballistic spin Hall current.\cite{schliemann,sinitsyn} Burkov and MacDonald%
\cite{burkov} recovered the universal ballistic value even in the dirty
limit, in which the broadening is larger than the SO-energy splitting. In
this Communication we study the effect of disorder on the spin Hall effect
in the diffuse regime, in which the scattering rate is larger than either
the frequency or the inverse sample traversal time, but for weak scattering.
By taking into account the vertex correction we find that the spin Hall
effect vanishes identically for short-range impurity scattering.

We model the disorder by randomly distributed isotropic short-range
potentials and compute longitudinal and transverse (Hall) conductivities for
both charge and spin currents by the Kubo formalism in the Born
approximation. The SO interaction is subject to a significant conductivity
vertex correction,\cite{inoue} which we find here to be decisive for the
spin Hall current. The vertex correction appears in such a way that the
current operator along the $x$-direction corresponding to the Rashba
Hamiltonian $J_{x}=e\left\{ (\hbar k/m)\mathbf{1}-\lambda \sigma
_{y}\right\} $ is modified by substituting $\lambda \rightarrow \tilde{%
\lambda}=\lambda +\lambda ^{\prime }$. Here $\sigma _{i}$ $(i=x,y,z)$ are
the Pauli spin matrices. The correction term $\lambda ^{\prime }$ is not
necessarily small compared with $\lambda $, and found to be $-\lambda $ in
the weak scattering regime. Only without the vertex correction, the spin
Hall conductivity tends to $e/8\pi $ as predicted by Sinova \textit{et al}..%
\cite{sinova} Physically, the diffuse scattering represented by the vertex
correction efficiently scrambles the precession of spins out of the 2DEG
plane induced by the applied electric field such that no net spin Hall
current remains. On the other hand, the induced spin accumulation in the $y-$%
direction is much less sensitive to impurity scattering.\cite{inoue} The
spin Hall conductivity may persist for long-range, anisotropic defect
potentials that correspond to predominantly forward scattering.

The Rashba Hamiltonian in the momentum representation and Pauli spin space
reads 
\begin{equation}
H_{0}=\left( 
\begin{array}{cc}
\frac{\hbar ^{2}}{2m}k^{2} & i\lambda \hbar k_{-} \\ 
-i\lambda \hbar k_{+} & \frac{\hbar ^{2}}{2m}k^{2}%
\end{array}%
\right) ,
\end{equation}%
where $k_{\pm }=k_{x}\pm ik_{y}\ $with $\mathbf{k=}\left( k_{x},k_{y}\right) 
$ the electron momentum in the 2DEG plane, and $\lambda \ $parametrizes the
tunable spin-orbit coupling. The eigenfunctions and eigenvalues of the
Hamiltonian corresponding to periodic boundary conditions are given as 
\begin{eqnarray}
\phi _{\mathbf{k}s} &=&\frac{1}{\sqrt{2L^{2}}}e^{i\mathbf{k}\cdot \mathbf{r}}%
\binom{is\frac{k_{-}}{k}}{1}, \\
E_{ks} &=&\frac{\hbar ^{2}k^{2}}{2m}+s\lambda \hbar k,
\end{eqnarray}%
respectively, with $s=\pm ,k=\sqrt{k_{x}^{2}+k_{y}^{2}}$ and $L^{2}$ the
area of the 2DEG. The corresponding (free) Green function is denoted as $%
g_{ks}(z)$ with an energy $z$ on the complex energy plane.

The disorder is modeled as randomly distributed, identical defects with
point scattering potentials that are neither spin-dependent nor flip the
spin: 
\begin{equation}
V(\mathbf{r})=V\mathbf{1}\sum_{i}\delta (\mathbf{r}-\mathbf{R}_{i}).
\end{equation}%
which gives rise to an isotropic $\left( s-\text{wave}\right) $ scattering
of electrons. The configurational averaged Green function reads%
\begin{equation}
\tilde{G}(k\pm )=\frac{1}{z-E_{k\pm }-\Sigma _{k\pm }(z)}.
\end{equation}%
In the Born approximation the self-energy $\Sigma (z)$ is a
state-independent constant: 
\begin{equation}
\Sigma (z)=\frac{nV^{2}}{2L^{2}}\sum_{ks}g_{ks}(z),
\end{equation}%
where $n$ is the impurity concentration and $L$ denotes the linear
dimensions of the sample. The self-energy is related to the scattering life
time $\tau $ via $|\mathrm{Im}\Sigma |=\hbar /2\tau $ at the Fermi energy $%
\epsilon _{F}$.

The charge current operator in spin space reads\cite{molenkamp} $%
J_{x}=e\partial H_{0}/\partial p_{x}=e\left( bk_{x}\mathbf{1}-\lambda \sigma
_{y}\right) $ and $J_{y}=e\partial H_{0}/\partial p_{y}=e\left( bk_{y}%
\mathbf{1}+\lambda \sigma _{x}\right) $ with $b=\hbar /m$. The spin currents
are represented by the Hermitian operators\cite{sinova}%
\begin{equation}
J_{\alpha }^{\sigma _{i}}=\dfrac{\hbar }{4}\left\{ \upsilon _{\alpha
},\sigma _{i}\right\} =\dfrac{\hbar }{4}\left\{ \dfrac{\partial H_{0}}{%
\partial p_{\alpha }},\sigma _{i}\right\} ,
\end{equation}%
where $\alpha =x,y,$ and $z$. Thus $J_{x}^{\sigma _{x}}=(\hbar
/2)bk_{x}\sigma _{x},J_{x}^{\sigma _{y}}=(\hbar /2)(-\lambda \mathbf{1}%
+bk_{x}\sigma _{y})$ and $J_{x}^{\sigma _{z}}=(\hbar /2)bk_{x}\sigma _{z}$,
whereas $J_{y}^{\sigma _{x}}=(\hbar /2)bk_{y}\sigma _{x},J_{y}^{\sigma
_{y}}=(\hbar /2)bk_{y}\sigma _{y},$ and $J_{y}^{\sigma _{z}}=(\hbar
/2)bk_{y}\sigma _{z}$.

The Kubo formula for the longitudinal electrical conductivity can be written 
\begin{equation}
\sigma _{xx}=\frac{\hbar }{2\pi L^{2}}\mathrm{Tr}\left\langle
J_{x}GJ_{x}G\right\rangle _{AV}.
\end{equation}%
where the trace is taken over wave vectors and band index. We evaluate $%
\langle J_{x}GJ_{x}G\rangle _{AV}=J_{x}\langle GJ_{x}G\rangle _{AV}\equiv
J_{x}K_{x}$ in the ladder approximation that obeys the Ward relation with
the self-energy in the Born approximation. This leads to the Bethe-Salpeter
equation 
\begin{equation}
K_{x}\approx \tilde{G}J_{x}\tilde{G}+\tilde{G}\langle VK_{x}V\rangle _{AV}%
\tilde{G}.
\end{equation}%
$K_{x}=\tilde{G}\tilde{J}_{x}\tilde{G}$ has the same structure as $\tilde{G}%
J_{x}\tilde{G}$, and\cite{inoue} 
\begin{equation}
\tilde{J}_{x}=e\left( bk_{x}\mathbf{1}+\dfrac{\tilde{\lambda}}{k}%
(k_{x}\sigma _{z}-k_{y}\sigma _{y})\right) .
\end{equation}%
with $\tilde{\lambda}=\lambda +\lambda ^{\prime }$. The vertex correction $%
\lambda ^{\prime }$ is the solution of%
\begin{equation}
\lambda ^{\prime }=\dfrac{nV^{2}}{4L^{2}}\sum_{k_{1}}\left[
bk_{1}G_{k_{1}}^{-}+(\lambda +\lambda ^{\prime })\left(
G_{k_{1}}^{+}+G_{k_{1}}^{+-}+G_{k_{1}}^{-+}\right) \right] ,  \label{vrtx1}
\end{equation}%
with $G_{k}^{s}=G_{k}^{++}+sG_{k}^{--}$ and $G_{k}^{ss^{\prime }}=\tilde{G}%
_{ks}\tilde{G}_{ks^{\prime }}$.

The generalized spin conductivity tensor in Pauli spin space reads%
\begin{equation}
\sigma _{\alpha x}^{\sigma _{i}}=\frac{\hbar }{2\pi L^{2}}\mathrm{Tr}%
J_{\alpha }^{\sigma _{i}}\langle GJ_{x}G\rangle _{AV}\sim \frac{\hbar }{2\pi
L^{2}}\mathrm{Tr}J_{\alpha }^{\sigma _{i}}K_{x},
\end{equation}%
where the vertex function is the same as before. Symmetry tells us that%
\begin{eqnarray}
\mathrm{Tr}J_{x}^{\sigma _{x}}K_{x} &=&\dfrac{e\hbar b^{2}}{8}\mathrm{Tr}%
k^{2}G^{+}\sigma _{x}+\dfrac{e\hbar b\tilde{\lambda}}{8}\mathrm{Tr}%
kG^{-}\sigma _{x}, \\
\mathrm{Tr}J_{x}^{\sigma _{y}}K_{x} &=&\dfrac{e\hbar b^{2}}{8}\mathrm{Tr}%
k^{2}G^{+}\sigma _{y}+\dfrac{e\hbar b(\tilde{\lambda}+\lambda )}{8}\mathrm{Tr%
}kG^{-}\sigma _{y}  \notag \\
&&-\dfrac{e\hbar b\lambda \tilde{\lambda}}{8}\mathrm{Tr}\left[ G^{+-}-G^{-+}%
\right] \sigma _{y}, \\
\mathrm{Tr}J_{x}^{\sigma _{z}}K_{x} &=&\dfrac{e\hbar b^{2}}{8}\mathrm{Tr}%
k^{2}G^{+}\sigma _{z}+\dfrac{e\hbar b\tilde{\lambda}}{8}\mathrm{Tr}%
kG^{-}\sigma _{z}, \\
\mathrm{Tr}J_{y}^{\sigma _{x}}K_{x} &=&\dfrac{e\hbar b\tilde{\lambda}}{8}%
\mathrm{Tr}k\left( G^{+-}-G^{-+}\right) \sigma _{y}, \\
\mathrm{Tr}J_{y}^{\sigma _{y}}K_{x} &=&-\dfrac{e\hbar b\tilde{\lambda}}{8}%
\mathrm{Tr}k\left( G^{+-}-G^{-+}\right) \sigma _{x}, \\
\mathrm{Tr}J_{y}^{\sigma _{z}}K_{x} &=&i\dfrac{e\hbar b\tilde{\lambda}}{8}%
\mathrm{Tr}k\left( G^{+-}-G^{-+}\right) \mathbf{1.}
\end{eqnarray}%
Because the Green functions depend only on $k$, the angular averages of $%
k_{x}^{2}$ and $k_{y}^{2}$ are $k^{2}/2$, and odd terms with respect $k_{x}$
and $k_{y}$ in the trace of the equations above vanish by symmetry. Without
SO interaction, all matrix elements of $J_{\alpha }^{\sigma _{i}}K_{x}$
vanish except for $J_{x}^{\sigma _{x}}K_{x}=J_{x}^{\sigma
_{y}}K_{x}=J_{x}^{\sigma _{z}}K_{x}$. But also these terms become zero after
taking the trace. This means that no spin current flows along the external
electric field.\cite{inoue} Only the spin Hall conductivity $\sigma
_{yx}^{\sigma _{z}}$ proportional to $\mathrm{Tr}J_{y}^{\sigma _{z}}K_{x}$
is nonzero, indicating that a spin Hall current along the $y-$direction and
polarized in the $z-$direction may exist when an external electric field is
applied along $x$, as predicted for the ballistic limit.\cite{sinova}

The magnitude of the spin Hall effect can be calculated easily by adopting
the following approximation for the product of Green functions at the Fermi
energy $\epsilon _{F}$, 
\begin{equation}
\tilde{G}(ks)\tilde{G}(ks)\approx \frac{2\pi \tau }{\hbar }\delta (\epsilon
_{F}-E_{ks}),  \label{apprx1}
\end{equation}%
which holds when the energy dependence of the self energy is weak and the
broadening is small compared to the SO energy splitting at the Fermi energy, 
$|\mathrm{Im}\Sigma |\ll 2\hbar \lambda k$. Then 
\begin{equation}
\sigma _{yx}^{\sigma _{z}}=-\sigma _{xy}^{\sigma _{z}}=\dfrac{e\tilde{\lambda%
}}{8\pi \lambda },
\end{equation}%
for $\epsilon _{F}>0$. This agrees with the ballistic result $\sigma
_{yx}^{\sigma _{z}}=e/8\pi $ by Sinova \textit{et al}.\cite{sinova} except
for a factor $\tilde{\lambda}/\lambda =1+\lambda ^{\prime }/\lambda $ due to
the vertex correction, but is identical to it when the vertex correction $%
\lambda ^{\prime }$ is neglected.

By substituting Eq. (\ref{apprx1}) into Eq. (\ref{vrtx1}), the vertex
correction $\lambda ^{\prime }$ is evaluated as $\lambda ^{\prime }=-\lambda 
$, \textit{i.e}. the spin Hall conductivity vanishes. Eq. (\ref{apprx1}) is
equivalent to the weak scattering or strong SO interaction limit. As far as
the spin Hall current is concerned, the effect of the impurity vertex
correction is thus found to be much more important than that of the impurity
self-energy in the Green function.

The ballistic result can be recovered by considering the frequency dependent
conductivity%
\begin{equation}
\sigma _{\mu \nu }^{\xi }=\lim_{\omega \rightarrow 0}\dfrac{Q_{\mu \nu
}^{\xi }(\omega )-Q_{\mu \nu }^{\xi }(0)}{-i\omega },
\end{equation}%
in terms of the correlation function%
\begin{eqnarray}
Q_{\mu \nu }^{\xi }(i\nu _{\ell }) &=&\dfrac{1}{L^{2}\beta }\sum_{m}\mathrm{%
Tr}\left[ J_{\mu }^{\xi }G(i\omega _{m}+i\nu _{\ell })J_{\nu }G(i\omega _{m})%
\right] \\
&=&\dfrac{1}{L^{2}}\mathrm{Tr}J_{\mu }^{\xi }K_{\nu }(i\nu _{\ell }),
\end{eqnarray}%
with%
\begin{equation}
K_{\nu }(i\nu _{\ell })=\dfrac{1}{\beta }\sum_{m}\langle G(i\omega _{m}+i\nu
_{\ell })J_{\nu }G(i\omega _{m})\rangle _{\mathrm{AV}}.
\end{equation}%
The vertex correction is calculated as before resulting in 
\begin{equation}
K_{x}(i\nu _{\ell })=\dfrac{1}{\beta }\sum_{m}\tilde{G}(i\omega _{m}+i\nu
_{\ell })\tilde{J}_{x}\tilde{G}(i\omega _{m})
\end{equation}%
where $\tilde{J}_{x}$ includes $\tilde{\lambda}=\lambda +\lambda ^{\prime
}(\omega )$ with 
\begin{equation}
\lambda ^{\prime }(\omega )=-\dfrac{\hbar }{\tau }\dfrac{\lambda }{-i\hbar
\omega +\dfrac{\hbar }{\tau }},
\end{equation}%
by letting $i\nu _{\ell }\rightarrow \hbar \omega +i0$. Here we invoked
again the weak scattering assumption.\cite{mahan} This result generalizes
Eq. (20). When the $\tau \rightarrow \infty $ limit is taken first, $\lambda
^{\prime }(\omega )\rightarrow 0,$ thus recovering the ballistic limit.\cite%
{sinova} When we take the $\omega \rightarrow 0$ limit first, $\lambda
^{\prime }(\omega )=-\lambda $, and the spin Hall conductivity vanishes as
before.

We made the rather crucial approximation that the scattering potential is
short-ranged, thus isotropic in momentum space. As mentioned above, Sinova 
\textit{et al}.\cite{sinova} explain the ballistic spin Hall current in
terms of the precession of spins out of the 2DEG plane when accelerated by
the electric field. In the presence of isotropic impurity scattering,
electrons with momentum $\mathbf{k}$ are scattered into all other momenta $%
\mathbf{k}^{\prime }$ at the Fermi energy with equal rate, and the spin Hall
current disappears with the average spin tilting. This picture is not
appropriate anymore when the impurity potentials are long-ranged, scattering
predominantly in the forward direction. In that case the short-range model
misrepresents the \textquotedblleft skew scattering\textquotedblright\
corresponding to a non-zero Hall angle.\cite{chazalviel}

For long-range anisotropic scatterers the longitudinal conductivity is
governed not by the energy life time $\tau $ but the transport (momentum)
life time $\tau _{t}$ because the momentum integration in the vertex
function over $V^{2}k_{x}$ ($x$ is the current direction) does not vanish.%
\cite{mahan} Physically this means that the forward (small angle) scattering
does not contribute to the resistivity. Without SO interaction, the vertex
correction due to anisotropic scattering reads%
\begin{equation}
b^{\prime }=\dfrac{\langle nV^{2}\rangle }{2L^{2}}\dfrac{1}{\kappa }%
\sum_{k_{1}}(b+b^{\prime })G_{k_{1}}^{+},  \label{vrtx2}
\end{equation}%
where $b=\hbar /m,$ $1/\kappa =\tau /\tau _{1}$, and $\langle nV^{2}\rangle $
is an average of the scattering potential over Fermi surface. The transport
life time is given by $1/\tau _{t}=1/\tau -1/\tau _{1}.$

When the SO interaction is incorporated into this vertex correction, the
expression of the longitudinal charge conductivity and spin accumulation
obtained before\cite{inoue} are modified as 
\begin{equation}
\sigma _{xx}=2\left[ \dfrac{e^{2}\tau _{t}}{m}n_{0}+e^{2}D\tau _{t}\lambda
^{2}\right] ,
\end{equation}%
and%
\begin{equation}
\langle s_{y}\rangle =2\tau _{t}eED\lambda ,
\end{equation}%
respectively. Here we have used following relations: $1/\tau =2\pi
nV^{2}D/\hbar =nV^{2}m/\hbar ^{3}$, with $D=m/2\pi \hbar ^{2}$, where $D$ is
the density of states of 2DEG. Note that the relation $\sigma
_{xx}^{\uparrow \uparrow }=\sigma _{xx}^{\downarrow \downarrow }$ holds for
arbitrary value of $b^{\prime }$ and $\lambda ^{\prime }$.

The spin Hall effect may survive when small angle scattering dominates
because only states close to each other in momentum space are scrambled. The
anisotropy may affect the effective current operator in Eq. (\ref{vrtx1}):
the first term in parenthesis on the right-hand side becomes $\lambda $ in
the isotropic scattering case and is likely to dominate for not too large
long-range potentials. The vertex correction $\lambda ^{\prime }$ is then
given by 
\begin{equation}
\lambda ^{\prime }=\dfrac{\langle nV^{2}\rangle }{4L^{2}}\dfrac{1}{\kappa
^{\prime }}\sum_{k_{1}}bG_{k_{1}}^{-},
\end{equation}%
in which $k_{1x}^{2}$ and $k_{1y}^{2}=k_{1}^{2}-k_{1x}^{2}$ are replaced
with weighted averages $k_{1}^{2}/2\kappa ^{\prime }$ and $%
k_{1}^{2}-k_{1}^{2}/2\kappa ^{\prime }$ over the angle. With $1/\kappa
^{\prime }=\tau /\tau ^{\prime }$, we get $\tilde{\lambda}=(\tau /\tau
_{H})\lambda $ and $1/\tau _{H}\equiv 1/\tau -1/\tau ^{\prime }$. In the
isotropic case, $\tau _{H}\rightarrow \infty $, and $\tilde{\lambda}%
\rightarrow 0$, but in general the spin-Hall current is finite. This
argument does not take into account the full effects of the anisotropy but
demonstrates how the vertex correction for anisotropic scattering affects
the spin-Hall conductivity.

Burkov and MacDonald\cite{burkov} computed the spin Hall conductivity for
the Rashba-2DEG model system with short-range impurity scattering, but
focusing on the dirty limit in which the lifetime broadening exceeds the SO
energy splitting, opposite to the clean limit discussed here. Surprisingly,
they recover the universal ballistic value found by Sinova \textit{et al}..%
\cite{sinova} This implies that with increasing (short-range) impurity
scattering the Hall conductivity first vanishes, but in a reentrant fashion
increases again when entering the dirty regime, in which Eq. (\ref{apprx1})
does not hold anymore.

Murakami \textit{et al}.\cite{murakami,murakami2} developed a theory for the
spin Hall currents in hole-doped semiconductors described by the Luttinger
Hamiltonian. Separating the spin Hall current into a topologically conserved
(intraband) and non-conserved (interband) contribution,\cite{murakami2}
these authors contend that the former, which does not exist in the
Rashba-2DEG, is robust against impurity scattering.\cite{sczhang} The
breakdown of the spin Hall current by impurity scattering in the Rashba-2DEG
discussed here would then correspond to the vanishing of the non-conserved
part of the spin-Hall current. Microscopic calculations for the Luttinger
Hamiltonian analogous to the present ones are necessary to unambiguously
prove that the topological spin Hall current indeed survives under impurity
scattering.

In conclusion, we have examined the effect of impurities on the spin Hall
conductivity of a Rashba-split 2DEG and found that the vertex correction
(diffuse electron scattering) to the conductivity is essential, causing the
spin-Hall effect to vanish.

The authors acknowledge fruitful discussions with Allan MacDonald, Jairo
Sinova, Shuichi Murakami, and Shoucheng Zhang. This work was supported by
the NEDO international project \textquotedblleft Nano-scale
Magnetoelectronics\textquotedblright , from Grants-in-Aid for Scientific
Research (C) and for Scientific Research in Priority Areas \textquotedblleft
Semiconductor Nanospintronics\textquotedblright\ of The Ministry of
Education, Culture, Sports, Science, and Technology of Japan, and by the FOM
Foundation, the DFG(SFB 410), and the DARPA Spins program.

\end{document}